\documentclass[sort&compress]{jfm}
\usepackage{graphicx}
\usepackage{amsmath, amssymb, esint}
\usepackage{color, hyperref, natbib}
\usepackage{lipsum}
\usepackage{tabularray}
\usepackage{gensymb}
\usepackage{soul,xcolor}
\setstcolor{red}

\newcommand{\no}{\noindent}

\def\XXint#1#2#3{{\setbox0=\hbox{$#1{#2#3}{\int}$}
     \vcenter{\hbox{$#2#3$}}\kern-.6\wd0}}

\title{Self-similarity and recurrence in stability spectra of near-extreme Stokes waves}

\author{B. Deconinck\aff{1}, S. A. Dyachenko\aff{2}, A. Semenova\aff{1}\corresp{\email{asemenov@uw.edu}}}

\affiliation{
	\aff{1}Department of Applied Mathematics, University of Washington
	\aff{2}Department of Mathematics, SUNY Buffalo.
}

\begin{document}

\maketitle

\begin{abstract}
We consider steady surface waves in an infinitely deep two--dimensional ideal fluid with potential flow, focusing on high-amplitude waves near the steepest wave with a 120$\degree$ corner at the crest. The stability of these solutions with respect to coperiodic and subharmonic perturbations is studied, using new matrix-free numerical methods. We provide evidence for a plethora of conjectures on the nature of the instabilities as the steepest wave is approached, especially with regards to the self-similar recurrence of the stability spectrum near the origin of the spectral plane.
\end{abstract}


\section{Introduction}

We consider the stability of spatially periodic waves that propagate with constant velocity without change of form, in potential flow of an ideal (incompressible and inviscid) two-dimensional fluid of infinite depth.
The study of such wave profiles has been the subject of much previous work, going back to~\cite{stokes1847theory} (also published in~\cite{stokes1880theory}). 
Stokes' work was followed by numerical computation of these waves by~\cite{michell1893}, and existence was proved in the works of~\cite{nekrasov1921waves} and~\cite{levi1925determination}, see also~\cite{toland1996stokes, hur2006global} for existence of the global branch of wave profiles.
The numerical study of these waves and the nature of their singularities was continued by~\cite{grant1973singularity},~\cite{schwartz1974computer},~\cite{williams1981limiting},~\cite{williams1985tables},~\cite{tanveer1993singularities},~\cite{cowley1999formation},~\cite{baker2011singularities} and others.

In the context of water waves, such waves are usually referred to as Stokes waves.
It was suggested by~\cite{stokes1880} that there exists a progressive wave of maximum height, and that
the angle at the crest of this limiting wave should be $2\pi/3$. Rigorous proofs of these Stokes conjectures came much later. 
\cite{toland1978existence} showed global existence of the limiting Stokes wave, but did not prove
that the angle at the crest is $2\pi/3$.
This result was proved by~\cite{amick1982stokes} and~\cite{plotnikov1982} (Reported in English in \cite{plotnikov2002proof}) independently. We refer to the Stokes wave of greatest height as the {\em extreme wave} and to waves of near-maximal amplitude as near-extreme waves. Even with the original Stokes conjectures resolved, the study of the graph of the wave profiles remains active, with a number of open problems, as detailed by \cite{dhs2023almost}, see also below.  The works by ~\cite{longuet1977theory,longuet1978theory} and by \cite{longuet2008approximation} study the near-extreme waves using both asymptotic and numerical methods. The review by \cite{haziotetcreview} discusses many currently active research directions.

The investigation of the stability of Stokes waves was begun in the works of~\cite{benjamin1967instability}, ~\cite{benjamin1967disintegration},~\cite{lighthill1965contributions} and~\cite{whitham1967non}. Except for the influential experimental work by~\cite{benjamin1967disintegration}, the focus of these works was on the dynamics of small disturbances of {\em small-amplitude} periodic Stokes waves. They unveiled the presence of the modulational or Benjamin-Feir instability with respect to long-wave disturbances in water of sufficient depth, $kh>1.363\ldots$. Here $h$ is the depth of the water and $k=2\pi/L$, with $L$ the period of the Stokes wave. The first rigorous results on the Benjamin-Feir instability were established by~\cite{bridgesmielke}, followed up very recently by~\cite{nguyenstrauss} and~\cite{hur2023unstable}. The numerical results of~\cite{deconinck2011instability} reveal the presence of a figure-8 curve in the complex plane of the spectrum of the linear operator governing the linear evolution of the Stokes wave disturbances. Approximations to this figure-8 are obtained by~\cite{creedon2023ahigh} and by~\cite{berti2022full}, where the existence of the figure-8 was proven rigorously. \cite{berti2023benjamin} also examined the critical case $kh=1.363\ldots$.

\cite{deconinck2011instability} also brought to the fore the presence of the so-called high-frequency instabilities, existing for narrow ranges of the disturbance quasi-periods. These instabilities were further studied by~\cite{creedon2022high} and by~\cite{hur2023unstable}, where their existence was proven rigorously. 

The instabilities mentioned above play a role in our study of the dynamics of 
large-amplitude Stokes waves, but we illustrate other instability mechanisms, not present for small-amplitude waves. Understandably, the study of large-amplitude Stokes waves, which cannot be thought of as perturbations of flat water, is harder, both from a computational and an analytical point of view. Nonetheless, some groundbreaking examinations have been done, for instance by ~\cite{tanaka1983stability},~\cite{longuet1997crest} and for near-extreme waves by~\cite{korotkevich2022superharmonic}. These authors all consider perturbations of the Stokes waves with respect to co-periodic (or superharmonic) disturbances, {i.e.}, the Stokes waves and the disturbance have the same minimal period. Their results are recapped in detail below, as they are instrumental to our own investigations. The results in this manuscript follow those of \cite{deconinck2023instability}, as we present a computational study of the instabilities of periodic Stokes waves, under the influence of disturbances parallel to the propagation direction of the wave. It should be emphasized that all figures presented below are 
quantitatively correct unless they are described as ``schematic'' in the caption. Similarly, all floating-point numbers given are approximate, of course, but all digits provided are believed to be correct.

\section{One-dimensional waves in water of infinite depth}

\begin{figure}
    \centering
    \includegraphics[width=0.99\textwidth]{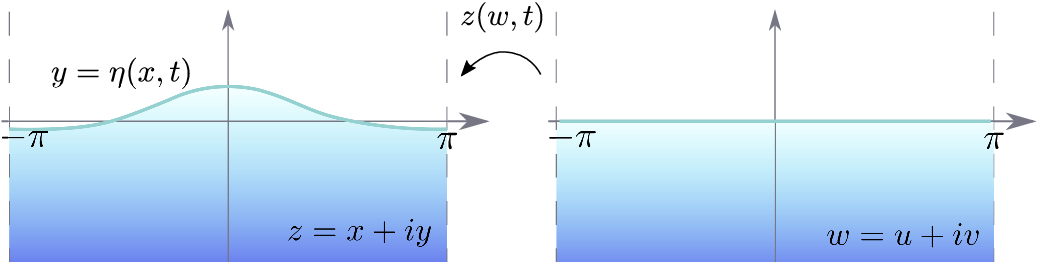}
    \caption{A schematic of the conformal map from the lower half plane (right) to the physical domain (left).}
    \label{fig:conf_map.eps}
\end{figure}

The equations of motion governing the dynamics of the one-dimensional free surface of a two-dimensional irrotational, inviscid fluid (see left panel of Fig.~\ref{fig:conf_map.eps}) are the Euler equations: 

\begin{align}
&&\Phi_{xx}+\Phi_{yy}=0, && x\in \mathbb{R},~y\in(-\infty, \eta(x,t)), \label{Laplace}\\
&&\eta_t= \left.\left(-\Phi_x \eta_x+\Phi_y\right)\right|_{y=\eta(x,t)}, && x\in \mathbb{R}, ~y=\eta(x,t),\label{kinematic_bc}\\
&&\left.\left(\Phi_t +\frac{1}{2} \left(\Phi_x^2+\Phi_y^2\right ) \right) 
\right|_{y = \eta(x,t)}+g\eta= 0, && x\in \mathbb{R}, ~y=\eta(x,t),\label{dynamic_bc}\\
&&\lim_{y\rightarrow -\infty} \Phi_y=0, && x\in \mathbb{R}. \label{deepbc}
\end{align}

\no Here $y=\eta(x,t)$ is the equation of the free surface and $\Phi(x,y,t)$ is the velocity potential (i.e., the velocity in the fluid is $v=(\Phi_x, \Phi_y)$), subscripts denote partial derivatives, $x$ and $y$ are the horizontal and vertical coordinate respectively, $t$ denotes time, and $g$ is the acceleration of gravity. We ignore the effects of surface tension. Although the Stokes waves are $2\pi$-periodic, it is important to pose the problem above on $x\in \mathbb{R}$, since the perturbations we consider are not necessarily periodic. The first equation expresses the divergence-free property of the flow under the free surface determined by $\eta(x,t)$. The second and third equation are nonlinear boundary conditions determining the free surface: the kinematic condition \eqref{kinematic_bc} expresses that the free surface changes in the direction of the normal derivative to the surface (particles on the surface remain on the surface), whereas the dynamic condition \eqref{dynamic_bc} states the continuity of pressure across the surface. Atmospheric pressure has been equated to zero, without loss of generality. 

Since the location of the surface $y=\eta(x,t)$ is the main focus of the water wave problem, different reformulations have been developed that eliminate the velocity potential in the bulk of the fluid as an unknown.  
\cite{zakharov1968} shows how the problem~\eqref{Laplace}--\eqref{deepbc} can be recast in terms of only the surface 
variables $\eta(x,t)$ and $\varphi(x,t)=\Phi(x,\eta(x,t),t)$, 
and the dynamics of $\eta(x,t)$ and $\varphi(x,t)$ is governed by an infinite-dimensional Hamiltonian system with $\eta(x,t)$ and $\varphi(x,t)$ as canonical variables. The Hamiltonian is the total energy of the system (with potential energy renormalized to account for infinite depth), which depends on the velocity potential $\Phi(x,y,t)$ in the bulk of the fluid. 

To avoid the dependence on the bulk, Zakharov's formulation uses the Dirichlet-to-Neumann operator (DNO), producing the normal derivative of the velocity potential at the free surface (the right-hand side of \eqref{kinematic_bc}) from the values of $\varphi(x,t)$. For small-amplitude waves, the DNO is conveniently expressed as a series, as done by \cite{CraigSulemJCompPhys1993}. For large-amplitude waves, such an expansion is not readily available, and the DNO has to be approximated numerically. To avoid doing so, 
we use conformal variables, see Fig.~\ref{fig:conf_map.eps}: for a $2\pi$-periodic wave, 
a time-dependent conformal transformation maps the half plane in the $w=u+i v$ plane ($(u,v) \in [-\pi,\pi]\times(-\infty,0]$)
into the area $(x,y) \in [-\pi,\pi]\times(-\infty,\eta]$ in the physical $z=x+iy$ plane occupied by the fluid. 
The horizontal line $v = 0$ is mapped into the fluid surface $y=\eta(x,t)$. The implicit equations of motion in conformal variables are constructed in the works of~\cite{ovsyannikov1973dynamika}, see also ~\cite{tanveer1991singularities},~\cite{dyachenko1996dynamics}, ~\cite{dyachenko2001dynamics}. 
We use this implicit formulation to study the stability of Stokes waves.

From these works, the conformal map $z(w,t) = x(w,t) + iy(w,t)$ is a complex-analytic function in $\mathbb{C}^-$ that approaches the identity map $z(w,t)\to w$ as $w \to -i\infty$, the image of a point at infinite depth. In the conformal variables, the Hamiltonian has the form

\begin{align}
    \mathcal{H} = \frac{1}{2}\int_{-\pi}^{\pi} \psi \hat k \psi \,du + \frac{g}{2}\int_{-\pi}^{\pi} y^2 x_u \,du,\label{HamConf}
\end{align}

\no where $\psi(u,t)=\varphi(x,t)$ and the operator $\hat k = -\hat H \partial_u$. Here $\hat H$ is the periodic Hilbert 
transform defined by the principal-value integral

\begin{align}
	\hat Hf(u)
= \frac{1}{2\pi} \fint^{\pi}_{-\pi} f(u') \cot{\frac{u'-u}{2}}du'. 
\end{align}

\no Equivalently, the Hilbert transform can be defined by its action on Fourier harmonics,
$\hat H e^{iku} = i\,\mbox{sign}(k)\, e^{iku}$.The equations of motion are derived by taking variational derivatives of the action $\mathcal{S} = \int \mathcal{L}\,dt$ with respect to $x$, $y$ and $\psi$. The Lagrangian has the form

\begin{align}
\mathcal{L} \!=\! \int_{-\pi}^\pi \!\!\!\psi \left(y_t x_u - y_u x_t\right)du \!-\! \mathcal {H} \!+\! \int_{-\pi}^\pi\!\!\! f(u)\left(y - \hat H \left[x-u\right]\right) du,
\end{align}

\no where the Lagrange multiplier $f(u)$ is chosen to enforce the relation $x(u,t) = u-\hat H \left[y(u,t)\right]$.
We refer to the work of \cite{DyachenkoEtAl1996} for the complete derivation of the equations of motion in conformal variables:

\begin{align}
y_t x_u - y_u x_t &= -\hat H\psi_u, \label{imp1}  \\
x_t \psi_u - x_u \psi_t - \hat H\left[y_t \psi_u -  y_u\psi_t \right] &=
g\left(x_uy - \frac{1}{2} \hat H \partial_u y^2\right). \label{imp2}
\end{align}

\subsection{Traveling Waves}

Using the conformal variables formulation (\ref{imp1}-\ref{imp2}), the Stokes waves are obtained by looking for a solution $y=y(u-ct)$, $\psi=\psi(u-ct)$, corresponding to stationary solutions in a frame of reference moving with constant speed $c$ in physical variables, see \cite{DyachenkoLushnikovKorotkevichJETPLett2014}. This gives rise to the so-called \cite{babenko1987some} equation:

\begin{align}
	\left(c^2 \hat k - g\right) y - \frac{g}{2}\left(\hat ky^2 + 2y\hat k y\right) = 0.\label{Babenko}
\end{align}

\no Since we are interested in the stability of near-extreme Stokes waves, the accurate numerical solution of \eqref{Babenko} for near-limiting values of the speed is required. Details of such computations for the Babenko equation~\eqref{Babenko} are given by~\cite{dyachenko2016branch}. In what follows, the ratio of the crest-to-trough height $H$ to wavelength $L$ is used as the definition of wave steepness $s = H/L$. 
The limiting Stokes wave has the steepness $s_{lim} = 0.1410634839\ldots$ and speed $c_{lim} = 1.0922850485\ldots$ as computed by~\cite{dhs2023almost}.

\begin{figure}
    \centering
    \includegraphics[width=1.0\textwidth]{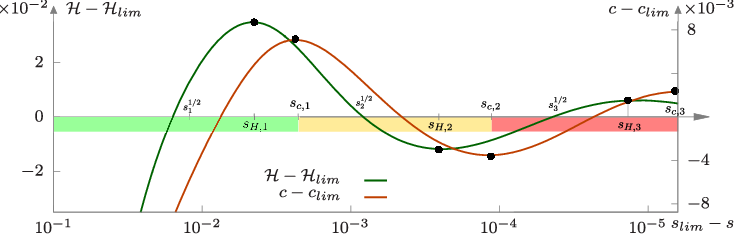}
    \caption{
    A schematic of the oscillations of the Hamiltonian ${\cal H}$ (green) and the velocity $c$ (red) relative to their limit values as steepness $s$ increases to its limiting 120$\degree$ Stokes wave value, using a logarithmic scale for the steepness $s$ relative to its limit value $s_{lim}$. In the green region, the waves are unstable with respect to the first Benjamin-Feir branch and the first localized instability branch, see Section~\ref{sec:instability}. In the yellow 
    region, they are unstable with respect to the second branches, and so on. 
    The steepnesses $s_{c,n}$ and $s_{H,n}$ correspond to the steepness values where the velocity $c$ and the Hamiltonian ${\cal H}$ have extreme values.
    }
    \label{fig:bestfig1}
\end{figure}

It is known that the speed $c$ and the Hamiltonian ${\cal H}$ oscillate as a function of the wave steepness $s$, for values near the limiting value $s_{lim}$, see~\cite{longuet1978theory}. It is believed there is an infinite number of such oscillations. They are presented schematically in Fig.~\ref{fig:bestfig1}. The other details of this figure constitute some of the main results of this paper, discussed below.

\subsection{Linearization about a Stokes wave}

In a reference frame traveling with the velocity of a Stokes wave, the Babenko equation describes a stationary solution of the equations~\eqref{imp1}--\eqref{imp2}. The linear stability of these Stokes waves is 
determined by the eigenvalue spectrum of the linearization in this traveling frame. To obtain the linearization, 
we transform (\ref{imp1}-\ref{imp2}) to a moving frame, using

\begin{align}
y(u,t) &\to y(u-ct,t), \quad \mbox{and} \quad x(u,t) \to u - \hat H y(u-ct,t), \label{ans1}\\ 
\psi(u,t) &\to\phi(u-ct, t) -  c\hat H y(u-ct,t), \label{ans2}
\end{align}

\no as in \cite{dyachenko2023canonical}. The nonlinear system~(\ref{imp1}--\ref{imp2}) in the moving frame becomes
\begin{align}
x_u y_t - y_u x_t &=  - \hat H\left(\phi_u - c \right), \label{KCm}\\ 
x_u \phi_t - x_t\phi_u - \hat H\left[y_u \phi_t - y_t \phi_u\right]
+ 2cx_t&= c^2\hat k y - g\left(x_uy - \hat H \left[yy_u\right]\right).\label{DCm}
\end{align}

\no The linearization about a Stokes wave is found by substituting

\begin{align}
y(u,t) &\to y(u) + \delta y(u,t)+\ldots  \quad \mbox{and}\quad  \phi(u,t) \to 0 + \delta \phi(u,t)+\ldots.
\end{align}

\no Here $y(u)$ corresponds to the Stokes wave and $\delta y(u,t)$, $\delta \phi(u,t)$ are small perturbations. 
Retaining only linear terms in $\delta y$ and $\delta \phi$ leads to the following evolution equations for the perturbations $\delta y$ and $\delta \phi$:

\begin{align}
\hat \Omega_{21}^\dagger \delta \phi_t - 2c\hat H \delta y_t &= \hat S_1 \delta y, \label{dcl}\\
\hat \Omega_{21}         \delta y_t &= \hat k \delta \phi,\label{kcl}
\end{align}

\no where the operator $\hat \Omega_{21}$ is defined via $\hat \Omega_{21} f = x_u f + y_u \hat H f$.  The operator $\hat \Omega^{\dagger}_{21}$ is its adjoint: $\hat \Omega^{\dagger}_{21} f = x_u f - \hat H\left[y_u f\right]$ and 

\begin{align}
\hat S_1(y)\delta y = \left(c^2\hat k - g\right)\delta y - 
g\left[y\hat k \delta y + \delta y \hat k y + \hat k (y\delta y) \right].\label{S1}
\end{align} 

\no This is rewritten in matrix form as

\begin{align}
&\hat L\partial_t
\left[
\begin{array}{c}
\delta \phi \\
\delta y
\end{array}
\right] =
\hat M
\left[
\begin{array}{c}
\delta \phi \\
\delta y
\end{array}
\right], \label{linsys1}
\end{align}

\no with
\begin{align}
\hat L =
\begin{bmatrix}
0  & \hat \Omega_{21} \\
\hat \Omega_{21}^\dagger  & -2c\hat H
\end{bmatrix},\,\, 
\hat M =
\begin{bmatrix}
\hat k  & 0 \\ 
 0  & \hat S_1
\end{bmatrix}.
\label{OperLin}
\end{align}

\no We examine the effect of quasi-periodic perturbations $\delta y$, $\delta \phi$ using the Fourier-Floquet-Hill (FFH) approach described in~\cite{deconinck2006computing} and \cite{deconinck2011instability}.  The time dependence for $\delta y$ and $\delta \phi$ is found using separation of variables. 
Moreover, in order to consider quasiperiodic perturbations we use a Floquet-Bloch decomposition in space, 

\begin{align}
\left[
\begin{array}{c}
\delta y(u,t)\\
\delta \phi(u,t)
\end{array}\right]=
e^{i\lambda t+i\mu u}\left[
\begin{array}{c}
\delta y(u)\\
\delta \phi(u)
\end{array}\right],
\end{align}

\no where  $\mu \in (-1/2,1/2]$ is the Floquet parameter and $\lambda(\mu) \in \mathbb{C}$  is the eigenvalue. 
The resulting $\mu$-dependent spectral problem is solved using a Krylov-based method and the shift-and-invert technique, see~\cite{dyachenko2023quasiperiodic}. Details on Krylov methods are presented by~\cite{stewart2002krylov}. We refer to the spectrum obtained this way as the stability spectrum of the Stokes wave. Note that the Floquet parameter is defined modulo 1, thus $\mu=1/2$ is equivalent to $\mu=-1/2$, see \cite{deconinck2006computing}.

\section{Instability}\label{sec:instability}

\subsection{The oscillating velocity and Hamiltonian}

Both the velocity $c$ and the Hamiltonian ${\cal H}$ depend on the Stokes wave. As the steepness $s$ of the Stokes wave increases and approaches its limiting value $s_{lim}$, both quantities are not monotone, as observed by \cite{longuet1975integral}. In fact, \cite{longuet1977theory,longuet1978theory} produce an asymptotic result that implies the presence of an infinite number of oscillations for both quantities. These oscillations were studied more by \cite{maklakov2002almost},~\cite{dyachenko2016branch} and~\cite{lushnikov2017new}, and very recently by \cite{dennisnew}. To our knowledge, no proof of an infinite number of oscillations in velocity $c$ and Hamiltonian ${\cal H}$ exists.

We denote the steepness 
of a wave at the $n$th turning point of the speed by $s_{c,n}$, $n=0, 1, 2, \ldots$, with $s_{c,0} = 0$. Similarly, the $n$th extremizer of the Hamiltonian is denoted by $s_{H,n}$. These critical values of the velocity and the Hamiltonian are important for changes in the stability spectrum, as shown below. For the Hamiltonian, the importance of these values is known, due to the works of \cite{tanaka1983stability,tanaka1985stability, saffman1985superharmonic, longuet1997crest}, for instance. It appears that these extremizing values interlace, so that $s_{c,n}<s_{H,n+1}<s_{c,n+1}$, $n=0, 1, \ldots$.  

In the recent work of \cite{dyachenko2023quasiperiodic}, 
a conjecture is made about an infinite number of secondary bifurcations associated with the Floquet multiplier $\mu=1/2$, corresponding to perturbations whose period is twice that of the Stokes wave. It is unclear how these bifurcations are related to those in the works of \cite{chen1980numerical}, \cite{longuet1985bifurcation} and \cite{zufiria1987non}, since those works do not introduce a Floquet parameter. However, their importance to the stability results presented here is demonstrated below. We denote the steepness associated with the $n$th secondary bifurcation point by $s_{n}^{1/2}$, $n=1,2, \ldots$. Further, we observe that $s_{c,n}<s_n^{1/2}<s_{H,n}$. These values are included in the schematic of Fig.~\ref{fig:bestfig1}.

It is convenient to break up the range of steepness $s$ from $s=0$ to $s=s_{lim}$ in intervals from one extremizer of the wave speed to the next. 
For example, the first interval starts at the primary bifurcation $s=0$ and ends at the first maximizer of the wave speed  $s=s_{c,1}$; the second interval starts at $s=s_{c,1}$ and ends at the first minimizer of the wave speed $s_{c,2}$, and so on. The length of each interval shrinks as the extreme wave is approached, and following \cite{longuet1978theory}
, we use a logarithmic scaling as illustrated in Fig.~\ref{fig:bestfig1}. Note that, because of the observed interlacing of the extremizers of the wave speed, those of the Hamiltonian, and the secondary bifurcations, each interval contains one extremizer of the Hamiltonian and one secondary bifurcation point. 

\subsection{A cyclus of changes in the spectrum}

\begin{figure*}
    \centering
    \includegraphics[width=0.95\textwidth]{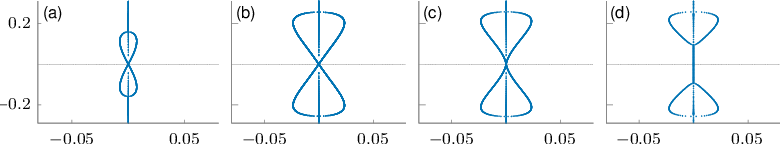}
    \includegraphics[width=0.95\textwidth]{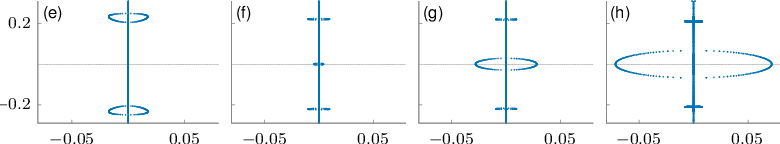}
    \includegraphics[width=0.95\textwidth]{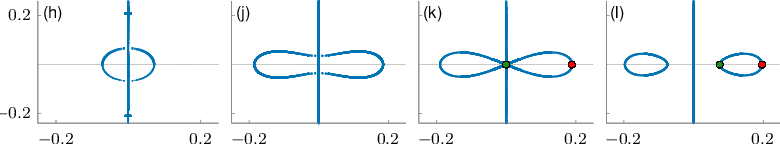}
    \caption{ Spectra in the vicinity of the origin for increasing steepness $s$. A detailed description is found in the main text. Here
    (a) $s = 0.0449032652$, (b) $s = 0.1042102092$, (c) $s = 0.1090618215$, (d) $s = 0.1122542820$,
    (e) $s = 0.1214481620$, (f) $s = 0.1289100582$, (g) $s = 0.1292029131$, (h) $s = 0.1307253066$,
    (j) $s = 0.1364173038$, (k) $s = 0.1366036552$, (l) $s = 0.1368557681$. Note that Panel (h) is repeated, with changing scales.}
    \label{fig:first_8}
\end{figure*}

As the steepness increases and each interval is traversed, an instability emerges from $\lambda = 0$ in the spectral plane, giving rise to a sequence of $\lambda(\mu)$-curves with changing topology, 
see Fig.~\ref{fig:first_8}. These changes for $s\in [s_{c,0}=0, s_{c,1})$ are described below.

\begin{figure*}
    \centering
    \includegraphics[width=0.95\textwidth]{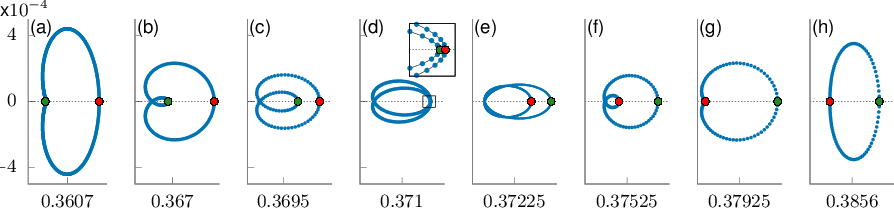}\\
    \includegraphics[width=0.95\textwidth]{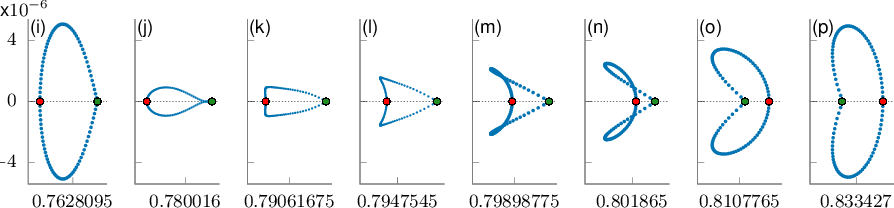}
    \caption{Spectrum component lobe with highest real part, for increasing steepness. A detailed description is found in the main text. Here (a) $s = 0.1394245282$, (b) $s = 0.1394647831$, (c)~$s = 0.1394802926$, (d) $s = 0.1394894509$, (e) $s = 0.1394970022$, (f) $s = 0.1395148411$, (g) $s = 0.1395380437$, (h) $s = 0.1395744737$, (i) $s = 0.1405658442$, (j) $s = 0.1405850778$, (k) $s = 0.1405964046$, (l) $s = 0.1406007221$, (m) $s = 0.1406050801$, (n) $s = 0.1406080087$, (o)~$s = 0.1406169126$, (p) $s = 0.1406384552$.}
    \label{fig:loc}
\end{figure*}

\begin{figure*}
    \centering
    \includegraphics[width=0.325\textwidth]{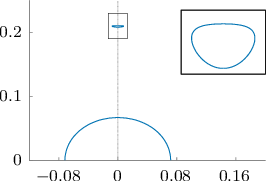}
    \includegraphics[width=0.325\textwidth]{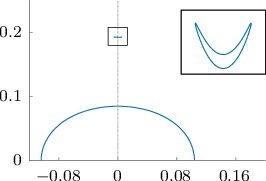}
    \includegraphics[width=0.325\textwidth]{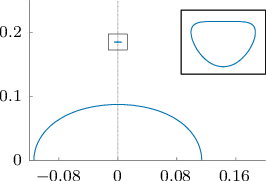}
    \caption{Changes in the Benjamin-Feir remnant as it approaches the oval at the origin, with steepness (from left to right panels): $s = 0.1307253066$, $s = 0.1323979204$, and $s = 0.1329490573$.}
    \label{fig:bf_peculiar}
\end{figure*}

\begin{enumerate}

\item Initially, at $s_{c,0}$, a figure-8 emerges from the origin, expanding in size as steepness increases, Panels (a) and (b). 

\item At an isolated value of the steepness $s=s_{h,1}$, both tangents of the figure-8 at the origin become vertical, resulting in an hourglass shape, Panel (c). 

\item Next, the lobes of the figure-8 detach from the origin, forming two disjoint isles qualitatively reminiscent of the high-frequency instabilities of~\cite{deconinck2011instability}. The band of Floquet values parameterizing the isles shrinks away from $\mu=0$ as the steepness increases, Panels (d)-(h). 

\item At $s=s_1^{1/2}$, eigenvalues with Floquet parameter $\mu=1/2$ bifurcate away from the origin onto the real line, creating an oval of eigenvalues with center at the origin, parameterized by Floquet values centered about $\mu=1/2$, see Panels (f)-(h). 

\item As the steepness increases, the oval around the origin deforms to a bean shape, eventually re-absorbing the remnants of the figure-8, Panel (j). More detail on the changes in these remnants and their re-absorption is presented in Figs.~\ref{fig:bf_peculiar} and \ref{fig:bf1decay}. 

\item At $s=s_{H,1}$, the bean shape pinches to form a figure-$\infty$. The double point of the figure-$\infty$ is at the origin and has a Floquet parameter $\mu=0$. Thus it corresponds to perturbations with the same period as the Stokes wave, see \cite{korotkevich2022superharmonic, dyachenko2023canonical}. In Panel (k) this co-periodic (or superharmonic) eigenvalue is marked in green. The unstable eigenvalue with $\mu=1/2$ is marked in red and gives rise to the most unstable mode for the wave with steepness $s=s_{H,1}$. 

\item As $s$ increases beyond $s_{H,1}$, the figure-$\infty$ splits off from the origin into a pair of symmetric lobes, one moving to the right, the other to the left, Panel (l). Further interesting changes in the shape of these lobes are observed as the steepness increases and they move away from the origin, see \cite{deconinck2023instability} and Fig.~\ref{fig:loc}. Importantly, we observe that the most unstable mode for this range of steepness $s$ is either co-periodic with the Stokes wave ($\mu=0$, green dot in Fig.~\ref{fig:loc}) or has twice its period, i.e., it is subharmonic with $\mu=1/2$ (red dot in Fig.~\ref{fig:loc}). Figure~\ref{fig:loc} illustrates two interchanges between these modes. We conjecture that such interchanges recur an infinite number of times as $s\rightarrow s_{lim}$. Note that the difference between $s_{lim}=0.1410634839$ and the steepness in the final panel of Fig.~\ref{fig:loc} is about $0.000425$ or $0.3\%$. 

\end{enumerate}

Below we focus on what happens near the origin of the spectral plane as the steepness continues to increase, ever getting closer to its extreme value.

\subsection{The Benjamin-Feir Instability}

We observe that a figure-8 shape in the stability spectrum emerges from the origin when the steepness $s=s_{c,n}$, an extremizer of the velocity. 
The first three extrema of the velocity appear at the following values:
\begin{align}
    s_{c,0} &= 0,\\   
    s_{c,1} &= 0.138753,\\ 
    s_{c,2} &= 0.140920. 
\end{align}

\no For small-amplitude waves (i.e., waves with steepness $s$ near $s_{c,0}=0$), the figure-8 corresponds to the well-studied classical Benjamin-Feir, or the modulational instability. In what follows we refer to instabilities manifested through a figure-8 in the spectral plane as Benjamin-Feir instabilities. 
We refer to the Benjamin-Feir instability branches starting at steepness $s_{c,n}$ as the $(n+1)$-th Benjamin-Feir branch, denoted BFI, BFII, BFIII, and so on. Below, we show that eigenvalues on the figure-8 near the origin (for BFII and BFII) give rise to modulational instabilities, as they do for small-amplitude waves, see \cite{benjamin1967instability, whitham1967non}.

\begin{figure}
    \centering
    \includegraphics[width=0.32\textwidth]{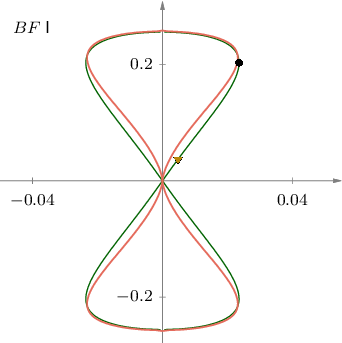}
    \includegraphics[width=0.32\textwidth]{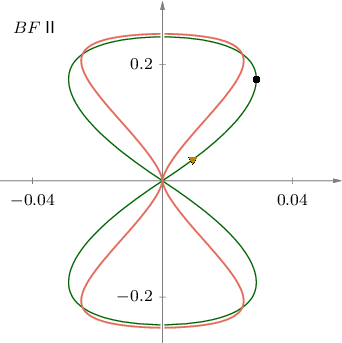}
    \includegraphics[width=0.32\textwidth]{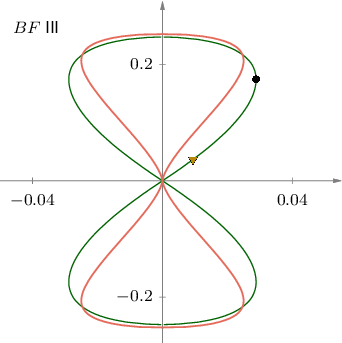}
    \caption{
    The figure-8 component of the eigenvalue spectrum, showing the Benjamin-Feir instability branches BFI, BFII and BFIII for Stokes waves of steepness 
    $s = 0.1045109822$ (left, green) and $s=0.1092129256$ (left, red);
    $s = 0.1398401087$ (center, green) and $s=0.1401021466$ (center, red); $s = 0.1409908317$ (right, green) and $s=0.1410079370$ (right, red). The green curves are associated with the maximal instability 
    growth on the corresponding Benjamin-Feir branch, with the black points marking the eigenvalues with largest real part, given in Table ~\ref{tab:bf123}. The red hourglass curves correspond to the steepness when the 
    figure-8 tangents at the origin become vertical, leading to the figure-8 detaching from the origin in the spectral plane. The points marked by gold triangles close to the origin have Floquet exponent $\mu=1/10$. 
    }
    \label{fig:origin}
\end{figure}

\begin{figure}
    \centering
    \includegraphics[width=0.495\textwidth]{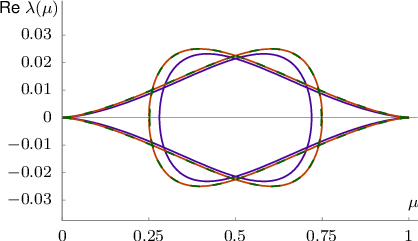}
    \includegraphics[width=0.495\textwidth]{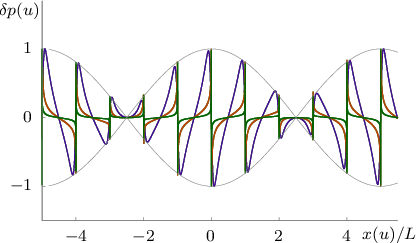}
    \caption{(Left Panel) The growth rate Re $\lambda$ as a function of the Floquet exponent $\mu$ for BFI, BFII and BFIII are shown in purple, dashed green and dashed red, respectively. The curves are associated with hourglass cases of the BFI, BFII and BFIII figure-8's, the red curves in Fig.~\ref{fig:origin}).
    (Right Panel) The unstable eigenfunctions associated with the triangular marker in Fig.~\ref{fig:origin}, left, center and right panel, are colored purple, red and green respectively. The envelope for all three is well described by $\cos(\mu x/L)$ with $\mu = 1/10$ and $L=2\pi$. Strong localization of the peaks of the eigenfunctions at the wave crests of the Stokes waves is observed as the wave steepness increases.
    } 
    \label{fig:origin2}
    \label{fig:bfefs}
\end{figure}

All the Benjamin-Feir branches that we compute experience the sequence of changes for increasing steepness  described above: they grow in size, their tangents at the origin become vertical followed by pinching off of the figure-8, resulting in the formation of isole on the positive and negative imaginary axis. For
each branch BFI, BFII and BFIII, we determine the figure-8 that gives rise to the eigenvalue with the largest real part, i.e., the maximal growth rate, see Fig.~\ref{fig:origin}. Table~\ref{tab:bf123} displays these values of steepness, the corresponding eigenvalue with maximal real part and its Floquet exponent, for BFI, BFII and BFIII.
These computations illustrate that the widest figure-8 (green curves in Fig.~\ref{fig:origin}) settles down to a universal shape as the extreme wave is approached, since the second and third panel appear indistinguishable). The values in Table ~\ref{tab:bf123} confirm this visual inspection. Further, for BFI, BFII and BFIII, we compute the hourglass shapes resulting from the figure-8's with vertical tangents at the origin, see Fig.~\ref{fig:origin}. We overlay these shapes in Fig.~\ref{fig:origin2}, plotting the real part of the spectrum as a function of the Floquet parameter. This figure illustrates convergence to a universal hourglass shape as the steepest wave is approached. We conjecture that an infinite number of Benjamin-Feir instability branches exist as the steepest wave is approached and that all of them experience a universal sequence of transitions.

Finally, using the points marked by triangles in Fig.~\ref{fig:origin}, we examine the eigenfunctions of \eqref{linsys1}. 
The eigenfunctions related to BFI, BFII and BFIII are visibly different, while their spectra in Figs.~\ref{fig:origin} and \ref{fig:origin2} are, to the eye, indistinguishable. A second observation is that these eigenfunctions are indeed modulational in nature: Fig.~\ref{fig:bfefs} displays $10~(=1/\mu)$ Stokes wave periods of the eigenfunctions. Although their effect is increasingly localized at the crest in each wave period, there is a more global modulational effect when many periods are considered. Thus even the BFII and BFIII instabilities, for $\mu$ close to zero, deserve the {\em modulational instability} moniker.

\begin{table}
  \begin{center}
\def~{\hphantom{0}}
  \begin{tabular}{l|ccc}
        & $s$ & $\lambda$ & $\mu$ \\\hline 
       BFI   & 0.1045109822 & 0.0235702 + 0.2029603i & 0.46376\\
       BFII   & 0.1398401087 & 0.0288896 + 0.1746076i & 0.45743\\
       BFIII   & 0.1409908317 & 0.0288299 + 0.1747554i & 0.45774\\
  \end{tabular}
  \caption{The Benjamin-Feir instability parameters for the figure-8 with the largest growth rate, first three branches.}
  \label{tab:bf123}    
  \end{center}
\end{table}

\subsection{The localized instability}

The $n$th oval emerges from the origin of the spectral plane as the steepness increases past $s^{1/2}_n$. We observe that $s^{1/2}_n>s_{h,n}$, the value of the steepness for which the $n$-th Benjamin-Feir figure-8 separates from the origin. Thus prior (i.e., for $s<s^{1/2}_n$) to these ovals emerging from the origin, the spectrum near the origin is confined to the imaginary axis. The primary, secondary, and tertiary ovals form at the steepnesses

\begin{align}
    s_{1}^{1/2} &= 0.128903,\\   
    s_{2}^{1/2} &= 0.140487,\\ 
    s_{3}^{1/2} &= 0.141032049, 
\end{align}
which correspond to steepnesses at which $4\pi$-periodic Stokes waves bifurcate from the primary, $2\pi$-periodic wave branch, see~\cite{dyachenko2023quasiperiodic}.

\begin{figure}
    \centering
    \includegraphics[width=0.495\textwidth]{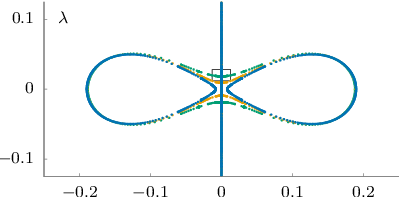}
    \includegraphics[width=0.495\textwidth]{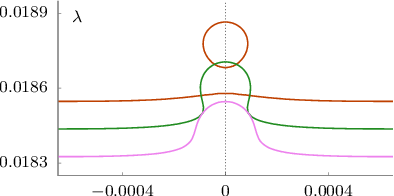}
    \caption{(Left) Bean-shaped stage of the localized instability for 
    $s = 0.1365552495 < s_{H,1}$ (green), $s = 0.1365917123 < s_{H,1}$ (gold) and 
    $s = 0.1366066477 > s_{H,1}$ (blue). (Right) Zoom-in on the remnant of the primary Benjamin-Feir isole as it is absorbed into the first localized branch at $s = 0.1365546598$ (red), $s = 0.1365552495$ (green) and $s = 0.1365558392$ (pink).}
    \label{fig:bf1decay}
\end{figure}

The changing topology of the primary oval for $s>s^{1/2}_1$ is shown in Figs.~\ref{fig:first_8} and \ref{fig:loc}.
More detail is presented in Fig.~\ref{fig:bf1decay}. The oval develops for $s_{1}^{1/2} \leq s < s_{H,1}$. The oval stage is followed by a symmetric bean shape with a narrowing neck as steepness approaches $s_{H,1}$. The maximal growth rate associated with the localized instability quickly overtakes the maximal growth rate associated with the Benjamin-Feir isole higher on the imaginary axis, see~\cite{deconinck2023instability}. Shortly before the steepness $s$  reaches $s_{H,1}$, the first extremizer of the Hamiltonian, the remnant of the Benjamin-Feir instability isola merges with the localized instability branch bean, as shown in the right panel of Fig.~\ref{fig:bf1decay}.

 We observe the recurrence of the process described above two more times, for the secondary and tertiary ovals that form at $s=s^{1/2}_2$ and $s=s^{1/2}_3$, respectively. This leads to the conjecture of an infinite number of such recurrences, the $n$-th one born at $s=s^{1/2}_n$, leading to the formation of the oval, gradually deforming to a bean shape, which pinches off at $s=s_{H,n}$, after which the resulting lobes move away from the origin along the real axis, ever decreasing in diameter. For $s>s_{H,n}$, the lobes are parameterized by the full range of Floquet exponents $\mu\in [-1/2, 1/2)$. Further, for $s\in (s_{H,n},s_{c,n+1})$ there is no component of the spectrum other than the imaginary axis. 

\begin{figure}
    \centering
    \includegraphics[width=0.495\textwidth]{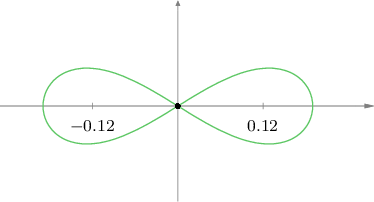}
    \includegraphics[width=0.495\textwidth]{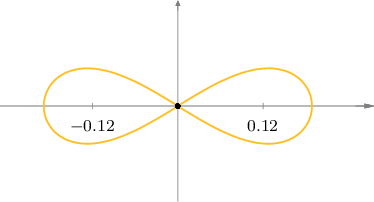}
    \caption{
    (Left) The figure-$\infty$ component of the instability spectrum appears at the first extremum of the Hamiltonian at $s_{H,1} = 0.1366035$.
    (Right) The second figure-$\infty$ component of the instability spectrum occurs for the wave with steepness $s_{H,2} = 0.1407965$ at the second extremum of the Hamiltonian. The difference between the real and imaginary parts of the two curves as a function of the Floquet exponent $\mu$ is less than $10^{-3}$.
    }
    \label{fig:infs}
\end{figure}
\begin{figure}
    \centering
    \includegraphics[scale=0.95]{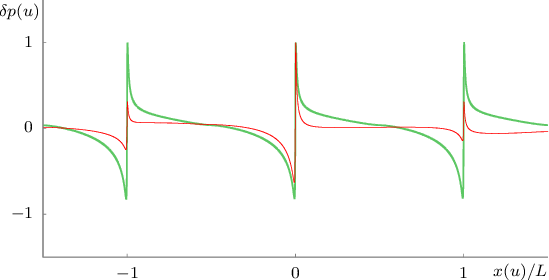}
    \includegraphics[scale=0.95]{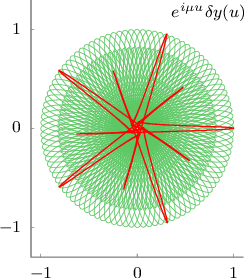}
    \caption{(Left) For the wave with $s_{H,1} = 0.1366035$ in the left panel of Fig.~\ref{fig:infs}, the perturbation associated with $\mu=0.01$ (green), with the eigenvalue $\lambda = 0.06106383 + 0.03263364i$ and its complex conjugate $\bar \lambda$. The perturbation is given by $\delta p = \mathrm{Re}\left[e^{i\mu u}\delta y\right]$. The red curve shows the same for $\mu=0.2$, with eigenvalue $\lambda = 0.1600750 + 0.04326491i$ and its complex conjugate $\bar \lambda$. Only the interval $-3\pi<x<3\pi$ is shown from the $2\pi/\mu$-periodic function. (Right) Polar plot, $e^{i\mu u}\delta y(u)$, 
    where real and imaginary parts are plotted along the horizontal and vertical axes respectively.}
    \label{fig:infs_eig}
\end{figure}
\begin{figure}
    \centering
    \includegraphics[scale=0.95]{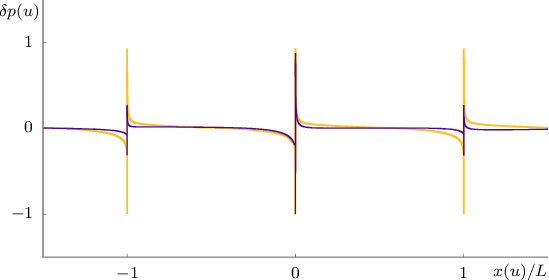}
    \includegraphics[scale=0.95]{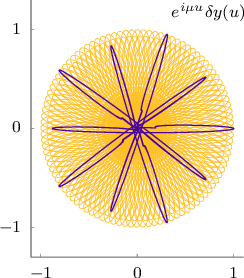}
    \caption{(Left) For the wave with $s_{H,2} = 0.1407965$ in the right panel of Fig.~\ref{fig:infs}, the perturbation associated with $\mu=0.01$ (gold) with the eigenvalue $\lambda = 0.06075090 + 0.03248890i$, and its complex conjugate $\bar \lambda$. The perturbation is given by $\delta p = \mathrm{Re}\left[e^{i\mu u}\delta y\right]$. The purple curve shows the same for $\mu=0.2$.  Only the interval $-3\pi<x<3\pi$ is shown from the $2\pi/\mu$-periodic function. (Right) Polar plot, $e^{i\mu u}\delta y(u)$, 
    where real and imaginary parts are plotted along the horizontal and vertical axes respectively.}
    \label{fig:infs_eig2}
\end{figure}

The first two figure-$\infty$'s are shows in Fig.~\ref{fig:infs}. Like the figure-$8$'s, they settle down to a universal shape as $s\rightarrow s_{lim}$. The difference between the real and imaginary parts of these first two figure-$\infty$'s as a function of the Floquet exponent never exceeds $10^{-3}$.

Some eigenfunctions associated with the figure-$\infty$ are displayed in Figs.~\ref{fig:infs_eig} and \ref{fig:infs_eig2}. For Floquet exponents close to zero (green and gold graphs), the modulational effect of the perturbation is clear from the polar plots. For other Floquet exponents (e.g. red and blue), the perturbation does not have a distinct modulational character. As for other high-amplitude Stokes waves, the localization of the eigenfunction (and thus the perturbation) near the crest of the waves is increasingly pronounced as $s\rightarrow s_{lim}$.

When the oval forms at $s=s_n^{1/2}$, its eigenvalue with largest real part is real and has Floquet exponent $\mu=1/2$, leading to eigenfunctions that have double the period of the Stokes wave. After pinch off, $s>s_{H,n}$, the left-most eigenvalue of the right lobe has $\mu=0$ (co-periodic eigenfunctions). As for the primary lobe, we conjecture that the most unstable mode on the right lobe is either the $\mu=0$ or the $\mu=1/2$ mode, which interchange an infinite number of times as $s\rightarrow s_{lim}$, see~\cite{deconinck2023instability}. As remarked above, the profile of the eigenfunction is strongly localized at the crests of the Stokes wave.  Modes with $\mu$ close to 0 have an envelope containing roughly $1/\mu$ periods of the Stokes wave and could be called modulational. However, in contrast to the Benjamin-Feir instabilities, for these instabilities the $\mu=0$ mode itself is unstable. For this reason, we refer to the instabilities emerging from the figure-$\infty$'s as {\em localized} instabilities. 

\begin{figure}
    \centering
    \includegraphics[width=0.99\textwidth]{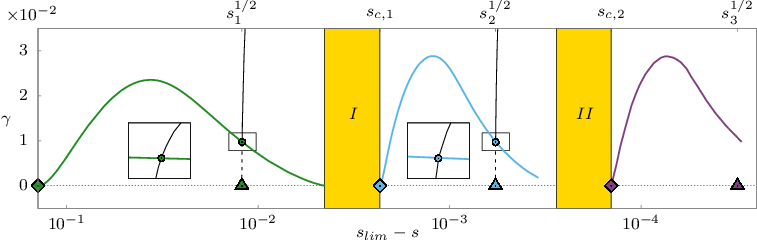}
    \caption{ The primary BF instability emerges at steepness $s=0$ for small-amplitude
    waves. This is marked with a green diamond. The green curve shows the maximal growth rate associated with this BF instability. At the secondary period-doubling bifurcation, $s_1^{1/2} = 0.128903$, marked with a green triangle, the first branch of the localized instability appears (maximal growth rate in black solid and dashed). The near-vertical appearance of this localized instability branch is a consequence of the rapid changes in the spectrum of this branch for steep waves. The inset $s \in [0.12878,0.12908]$, $\gamma \in [0.005,0.018]$ shows a zoom-in of the intersection of the maximal growth rate of the localized branch with the primary BF branch. The remnant of the BF instability merges with the localized branch at the edge of the gold region I. In this region, the BF branch is no longer distinguishable from the localized branch. The secondary branch of BF (maximal growth rate plotted in blue) emerges at the first maximizer of the speed at $s_{c,1} = 0.138753$ (blue diamond) and follows the same sequence of steps merging with the localized branch at the edge of the golden rectangle II. The secondary localized branch emerges at the second period-doubling bifurcation at $s_2^{1/2} = 0.140487$ (blue triangle). For the secondary inset $s \in [0.1404795, 0.1404955], \gamma \in [0.005,0.018]$. 
    The tertiary BF emerges at the turning point of speed, $s_{c,2} = 0.140920$ (maximal growth rate plotted in purple), the tertiary localized branch appears at $s_3^{1/2} = 0.141032049$ (purple triangle). 
    }
    \label{fig:bf12_growth}
\end{figure}

\subsection{The maximal growth rate} 

We track the maximum growth rate $\gamma$ (i.e., we track the eigenvalues with the largest real part) of the Benjamin-Feir and localized instabilities (plotted in black dotted and solid lines) as a function of the steepness of the Stokes wave in Fig. ~\ref{fig:bf12_growth}. The maximum growth rate for BFI, BFII, and BFII are presented by green, blue, and purple curves respectively. 
Steepnesses at which the dominant instability switches from the Benjamin-Feir to the corresponding localized branch are marked by circles. These switches are presented in the corresponding insets.
The steepness values where the maximal growth rates for BFI and BFII vanish, correspond to the case where the Benjamin-Feir remants are absorbed into the localized instabilities.

\subsection{The high-frequency instabilities} 

\begin{figure}
    \centering
    \includegraphics[width=0.495\textwidth]{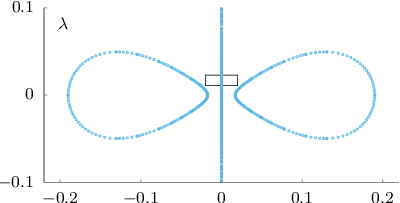}
    \includegraphics[width=0.495\textwidth]{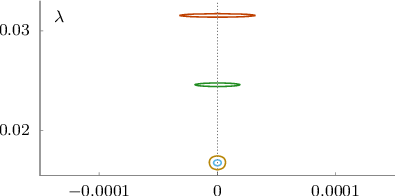}
    \caption{(Left) Detached figure-$\infty$ spectrum of the localized instability for $s = 0.1366171604 > s_{H,1}$ (blue). (Right) Zoom-in on the high-frequency instability for four increasing values of the steepness, $s = 0.1366066477$ (red), $s = 0.1366141405$ (green), $s = 0.1366171424$ (golden), and $s = 0.1366171604$ (blue, the value for the left panel), showing the vanishing of a high-frequency isola at a value just exceeding $s=s_{H,1}$.}
    \label{fig:bf1decay2}
\end{figure}

Since the work presented here focuses on the evolution of the spectrum for increasing steepness in the vicinity of the origin in the spectral plane, we discuss the high-frequency instabilities only briefly. As shown by \cite{deconinck2011instability} and \cite{creedon2022high}, the high-frequency instabilities emanate from purely imaginary double eigenvalues for steepness $s=0$, giving rise to an isola of unstable eigenvalues centered on the imaginary axis, away from the origin. As steepness is increased, these isole may collapse back on the imaginary axis, and new ones may form, see \cite{mackay1986stability}. Unlike the Benjamin-Feir (figure-8's) and localized instability branches (figure-$\infty$'s), the high-frequency isole are highly localized in the space of Floquet exponents: indeed, \cite{deconinck2011instability} show that often a range of Floquet exponents of width no more than $10^{-4}$ parameterizes an isola. This complicates their numerical detection. Figure~\ref{fig:bf1decay2} presents plots of a few high-frequency isole for near-extreme increasing steepness, showing the collapse of one into the imaginary axis. For the top isola plotted, 
$\mu\in [0.00092,0.00095]$, for the one below $\mu\in [0.00057,0.00059]$. For the two isole on bottom, $\mu\in [0.000251, 0.000294]$ (outer), $\mu\in [0.000263,0.000282]$ (inner). This demonstrates the isole can be captured using our method. A detailed study of the evolution of the high-frequency instabilities as steepness increases is kept for future work. 

\section{Conclusions}

\begin{figure}
    \centering
    \includegraphics[width=1.0\textwidth]{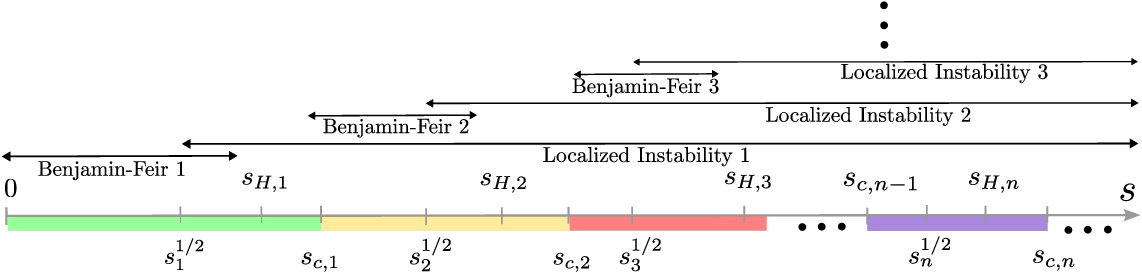}
    \caption{A schematic view of the appearance (and existence for a range of steepnesses) of localized and Benjamin-Feir type instabilities. The first Benjamin-Feir instability figure-8 appears at the steepness $s = s_{c,0}=0$, the second one appears at $s=s_{c,1}$, and the third one at $s=s_{c,2}$. Localized instabilities, manifested by an oval at the origin deforming to a figure-$\infty$ appear at steepnesses labeled $s^{1/2}_n$ with $n=1,2,3, ...$ that correspond to bifurcations to double-period Stokes waves. Once localized instabilities appear, they continue to exist for all larger values of the steepness in contrast to Benjamin-Feir type instabilities that emerge and vanish as the steepness is increased. We conjecture that infinitely many Benjamin-Feir type and localized instabilities appear as $s\rightarrow s_{lim}$, the steepness of the extreme wave.
    }
    \label{fig:bestfig2}
\end{figure}

We have presented a numerical exploration of the stability spectrum of Stokes waves near the origin of the spectral plane, focusing on the topological changes in the stability spectrum as the wave steepness grows. The main challenge of the study is due to the non-smooth nature of the extreme Stokes wave, which has a 120$\degree$ corner at its crest. Thus for waves whose steepness approaches that of this extreme wave, our Fourier-based method requires the use of millions of Fourier modes. Indeed, for the computation of waves with steepness near the third extremizer of the Hamiltonian (see Fig.~\ref{fig:bestfig1}) nearly ten million Fourier modes are used. To examine the stability of these waves, we linearize about them, resulting in a generalized operator eigenvalue problem. The numerical approximation of this problem results in a generalized matrix eigenvalue problem with matrices of dimension equal to the square of twice the number of Fourier modes used for approximating the underlying Stokes wave, since each component of the perturbation requires a comparable number of Fourier modes to reach the same accuracy. Storing and manipulating such matrices is prohibitive, and our investigations are only possible because of the matrix-free approaches to the conformal mapping formulation, introduced by~\cite{dyachenko2023canonical}. 

In \cite{deconinck2023instability} we used this same method to investigate the largest growth rate of perturbations of near-extreme Stokes waves, as a function of their steepness. Among other outcomes, this led to the conclusion that long-lived ocean swell is confined to moderate amplitudes. In this paper, we focus instead on the behavior of the stability spectrum near the origin of the spectral plane, as the recurring, self-similar behavior may provide an indication of how the stability of near-extreme Stokes waves may be approached analytically. Specifically, previous work and our numerical explorations lead to the following conjectures. 

\begin{enumerate}

\item The Hamiltonian ${\cal H}$ and the velocity ${c}$ have an infinite number of local extrema ($s_{H,n}$, $n=1,2,\ldots$ and $s_{c,n}$, $n=0,1,\ldots$ respectively) as the steepness $s$ increases to $s_{lim}$, the steepness of the extreme wave. This conjecture is not new, and the asymptotics of~\cite{longuet1977theory, longuet1978theory} provides a strong indication as to its validity. We include this conjecture here because all others below depend upon it. 

\item The maximal instability growth rate approaches infinity as the steepness increases to that of the steepest wave. Convincing evidence for this conjecture is presented by~\cite{korotkevich2022superharmonic}. This would imply that the Euler water wave problem for the evolution of the extreme Stokes wave is ill posed. This is not a surprise as capillary effects need to be incorporated when the curvature at the crest is too large. This is discussed in more detail by~\cite{deconinck2023instability}. 

\end{enumerate}

The conjectures below are a direct outcome of the investigations presented in this paper. A graphical overview of which features occur at which steepness, according to these conjectures, is presented in Fig.~\ref{fig:bestfig2}.

\begin{enumerate}

\item[(iii)] As the steepness $s$ increases from 0 to that of the steepest wave, there exists an infinite number of Benjamin-Feir figure-8 instabilities. These emanate from the origin at each extremum $s_{c,n}$, $n=0,1, \ldots$ of the velocity. Upon formation, these figure-8's persist for a range of steepness. After their tangents at the origin become vertical (resulting in an hourglass shape) at steepness $s_{h,n}$, $n=1,2,\ldots$, the figure-8 separates in two lobes on the imaginary axis. 

\item[(iv)] After the Benjamin-Feir figure-8 separates from the origin, the $n$th oval appears at the origin, for the steepness $s^{1/2}_n$, $n=1,2,\dots$, 
corresponding to the period-doubling bifurcation points from the primary branch of Stokes waves.

\item[(v)] As the steepness $s$ increases from 0 to that of the extreme wave, there are an infinite number of figure-$\infty$ instabilities. These occur at the origin at each extremum of the Hamiltonian/energy. As the steepness is increased, the figure-$\infty$'s detach instantaneously. In other words, the figure-$\infty$ shapes occur only at isolated values of the steepness, for which the Hamiltonian has a local extremum. 

\item[(vi)] These figure-8's and figure-$\infty$'s alternate in their occurrence. Stated differently, the extremizers of the Hamiltonian interlace the extremizers of the velocity. 

\item[(vii)] For Stokes waves with amplitude greater than that of the first maximizer of the Hamiltonian, the most unstable mode is either superharmonic (co-periodic) or subharmonic with twice the period of the Stokes wave (anti-periodic). Further, there exists an infinite number of interchanges between which of these two modes is dominant. Throughout these interchanges, no other mode is the most unstable.
These observations were already discussed  by~\cite{deconinck2023instability}.

\end{enumerate}

In the context of the full water wave problem, the present work reveals two primary mechanisms for the breaking of ocean waves: (i) for Stokes waves with $s < s_2=0.12894$ (the steepness at which the dominant instability switches from Benjamin-Feir to the localized branch, the abscissa of the green circle in Fig.~\ref{fig:bf12_growth}), when the unstable envelope of multiple periods of a train of Stokes waves enters the nonlinear stage, a complicated pattern of waves is observed on the free surface. The pattern tends to self focus leading to the formation of a large 
unsteady wave whose crest forms a plunging breaker~\cite{clamond2006long,onorato2013rogue}; 
(ii) for Stokes waves with $s > s_2$, the dynamics 
of the wave 
is dominated by the localized instability at the wave crest, see~\cite{dyachenko2016whitecapping, baker2011singularities, duncan2001spilling}. The localized instability immediately leads to wave breaking of either 
every other wave crest in the train (if $\mu = 0$ is dominant), or every other
crest (if $\mu = 1/2$ is dominant) as discussed by~\cite{deconinck2023instability}. More work is needed to understand the fully nonlinear stage of the many different instabilities computed. 

A complete understanding of the stability of Stokes waves with respect to bounded perturbations (see~\cite{haragucs2008spectra, kapitulapromislow}) requires further study of the spectrum of the operators associated with the linearization of the Euler equations governing the dynamics of these perturbed waves. Nonetheless, our study provides numerical evidence that the (quasi-) periodic eigenfunctions that we examine, are fundamental to this problem. 

\vspace*{0.1in}

\noindent {\bf Acknowledgements.} The authors wish to thank Eleanor Byrnes, Diane M. Henderson, and Pavel M. Lushnikov for helpful discussions. Also, the authors thank~\cite{frigo2005design}, the developers of FFTW and the whole GNU project for developing and supporting this important and free software. S.D. thanks the Isaac Newton Institute for Mathematical Sciences, Cambridge, UK, for support and hospitality during the programme ``Dispersive hydrodynamics'' where work on this paper was partially undertaken.  Partial support for A. S. is provided by a PIMS-Simons postdoctoral fellowship.

\vspace*{0.1in}

\noindent {\bf Declaration of Interests.} The authors report no conflict of interest.

\bibliographystyle{jfm}
\bibliography{water-waves-refs}

\begin{thebibliography}{71}
\expandafter\ifx\csname natexlab\endcsname\relax\def\natexlab#1{#1}\fi
\def\au#1{#1} \def\ed#1{#1} \def\yr#1{#1}\def\at#1{#1}\def\jt#1{\textit{#1}}
  \def\bt#1{#1}\def\bvol#1{\textbf{#1}} \def\vol#1{#1} \def\pg#1{#1}
  \def\publ#1{#1}\def\arxiv#1{#1}\def\org#1{#1}\def\st#1{\textit{#1}}

\bibitem[Amick {\em et~al.\/}(1982)Amick, Fraenkel \& Toland]{amick1982stokes}
{\sc \au{Amick, C.J.}, \au{Fraenkel, L.E.} \& \au{Toland, J.F.}} \yr{1982}
  \at{{On the Stokes conjecture for the wave of extreme form}}.  \jt{Acta
  Mathematica}  \bvol{148}~(1),  \pg{193--214}.

\bibitem[Babenko(1987)]{babenko1987some}
{\sc \au{Babenko, K.I.}} \yr{1987} Some remarks on the theory of surface waves
  of finite amplitude.  \bt{In {\em Doklady Akademii Nauk\/}}, ,  \vol{vol.
  294},  \pg{pp. 1033--1037}. Russian Academy of Sciences.

\bibitem[Baker \& Xie(2011)]{baker2011singularities}
{\sc \au{Baker, G.R.} \& \au{Xie, C.}} \yr{2011}  \at{{Singularities in the
  complex physical plane for deep water waves}}.  \jt{{Journal of Fluid
  Mechanics}}  \bvol{685},  \pg{83--116}.

\bibitem[Benjamin(1967)]{benjamin1967instability}
{\sc \au{Benjamin, T.B.}} \yr{1967}  \at{{Instability of periodic wavetrains in
  nonlinear dispersive systems}}.  \jt{{Proceedings of the Royal Society of
  London. Series A. Mathematical and Physical Sciences}}  \bvol{299}~(1456),
  \pg{59--76}.

\bibitem[Benjamin \& Feir(1967)]{benjamin1967disintegration}
{\sc \au{Benjamin, T.B.} \& \au{Feir, J.E.}} \yr{1967}  \at{{The disintegration
  of wave trains on deep water}}.  \jt{{Journal of Fluid Mechanics}}
  \bvol{27}~(3),  \pg{417--430}.

\bibitem[Berti {\em et~al.\/}(2022)Berti, Maspero \& Ventura]{berti2022full}
{\sc \au{Berti, M.}, \au{Maspero, A.} \& \au{Ventura, P.}} \yr{2022}  \at{{Full
  description of Benjamin-Feir instability of Stokes waves in deep water}}.
  \jt{{Inventiones mathematicae}}  \bvol{230}~(2),  \pg{651--711}.

\bibitem[Berti {\em et~al.\/}(2023)Berti, Maspero \&
  Ventura]{berti2023benjamin}
{\sc \au{Berti, M.}, \au{Maspero, A.} \& \au{Ventura, P.}} \yr{2023}
  \at{{Benjamin--Feir Instability of Stokes Waves in Finite Depth}}.
  \jt{{Archive for Rational Mechanics and Analysis}}  \bvol{247}~(5),  \pg{91}.

\bibitem[Bridges \& Mielke(1995)]{bridgesmielke}
{\sc \au{Bridges, T.} \& \au{Mielke, A.}} \yr{1995}  \at{{A proof of the
  Benjamin-Feir instability}}.  \jt{{Archive for Rational Mechanics and
  Analysis}}  \bvol{133},  \pg{145--198}.

\bibitem[Chen \& Saffman(1980)]{chen1980numerical}
{\sc \au{Chen, Benito} \& \au{Saffman, PG}} \yr{1980}  \at{Numerical evidence
  for the existence of new types of gravity waves of permanent form on deep
  water}.  \jt{Studies in Applied Mathematics}  \bvol{62}~(1),  \pg{1--21}.

\bibitem[Clamond {\em et~al.\/}(2006)Clamond, Francius, Grue \&
  Kharif]{clamond2006long}
{\sc \au{Clamond, D.}, \au{Francius, M.}, \au{Grue, J.} \& \au{Kharif, C.}}
  \yr{2006}  \at{Long time interaction of envelope solitons and freak wave
  formations}.  \jt{European Journal of Mechanics-B/Fluids}  \bvol{25}~(5),
  \pg{536--553}.

\bibitem[Cowley {\em et~al.\/}(1999)Cowley, Baker \&
  Tanveer]{cowley1999formation}
{\sc \au{Cowley, S.J.}, \au{Baker, G.R.} \& \au{Tanveer, S.}} \yr{1999}
  \at{{On the formation of Moore curvature singularities in vortex sheets}}.
  \jt{{Journal of Fluid Mechanics}}  \bvol{378},  \pg{233--267}.

\bibitem[Craig \& Sulem(1993)]{CraigSulemJCompPhys1993}
{\sc \au{Craig, W.} \& \au{Sulem, C.}} \yr{1993}  \at{{Numerical simulation of
  gravity waves}}.  \jt{{Journal of Computational Physics}}  \bvol{108},
  \pg{73--83}.

\bibitem[Creedon \& Deconinck(2023)]{creedon2023ahigh}
{\sc \au{Creedon, R.} \& \au{Deconinck, B.}} \yr{2023}  \at{{A high-order
  asymptotic analysis of the Benjamin--Feir instability spectrum in arbitrary
  depth}}.  \jt{{Journal of Fluid Mechanics}}  \bvol{956},  \pg{A29}.

\bibitem[Creedon {\em et~al.\/}(2022)Creedon, Deconinck \&
  Trichtchenko]{creedon2022high}
{\sc \au{Creedon, R.P.}, \au{Deconinck, B.} \& \au{Trichtchenko, O.}} \yr{2022}
   \at{{High-frequency instabilities of Stokes waves}}.  \jt{Journal of Fluid
  Mechanics}  \bvol{937}.

\bibitem[Deconinck {\em et~al.\/}(2023)Deconinck, Dyachenko, Lushnikov \&
  Semenova]{deconinck2023instability}
{\sc \au{Deconinck, B.}, \au{Dyachenko, S.A.}, \au{Lushnikov, P.M.} \&
  \au{Semenova, A.}} \yr{2023}  \at{{The dominant instability of near-extreme
  Stokes waves}}.  \jt{{Proceedings of the National Academy of Sciences}}
  \bvol{120}~(32),  \pg{e2308935120}.

\bibitem[Deconinck \& Kutz(2006)]{deconinck2006computing}
{\sc \au{Deconinck, B.} \& \au{Kutz, J.N.}} \yr{2006}  \at{Computing spectra of
  linear operators using the {F}loquet--{F}ourier--{H}ill method}.
  \jt{{Journal of Computational Physics}}  \bvol{219}~(1),  \pg{296--321}.

\bibitem[Deconinck \& Oliveras(2011)]{deconinck2011instability}
{\sc \au{Deconinck, B.} \& \au{Oliveras, K.}} \yr{2011}  \at{{The instability
  of periodic surface gravity waves}}.  \jt{{Journal of Fluid Mechanics}}
  \bvol{675},  \pg{141--167}.

\bibitem[Duncan(2001)]{duncan2001spilling}
{\sc \au{Duncan, J.H.}} \yr{2001}  \at{Spilling breakers}.  \jt{{Annual Review
  of Fluid Mechanics}}  \bvol{33},  \pg{519}.

\bibitem[Dyachenko(2001)]{dyachenko2001dynamics}
{\sc \au{Dyachenko, A.I.}} \yr{2001} {On the dynamics of an ideal fluid with a
  free surface}.  \bt{In {\em {Doklady Mathematics}\/}}, ,  \vol{vol.~63},
  \pg{pp. 115--117}. Pleiades Publishing, Ltd.

\bibitem[Dyachenko {\em et~al.\/}(1996)Dyachenko, Kuznetsov, Spector \&
  Zakharov]{DyachenkoEtAl1996}
{\sc \au{Dyachenko, A.I.}, \au{Kuznetsov, E.A.}, \au{Spector, M.D.} \&
  \au{Zakharov, V.E.}} \yr{1996}  \at{{Analytical description of the free
  surface dynamics of an ideal fluid (canonical formalism and conformal
  mapping)}}.  \jt{{Physics Letters A}}  \bvol{221}~(1-2),  \pg{73--79}.

\bibitem[Dyachenko {\em et~al.\/}(2023)Dyachenko, Hur \&
  Silantyev]{dhs2023almost}
{\sc \au{Dyachenko, S.A.}, \au{Hur, V.M.} \& \au{Silantyev, D.A.}} \yr{2023}
  \at{{Almost extreme waves}}.  \jt{{Journal of Fluid Mechanics}}  \bvol{955},
  \pg{A17}.

\bibitem[Dyachenko {\em et~al.\/}(2014)Dyachenko, Lushnikov \&
  Korotkevich]{DyachenkoLushnikovKorotkevichJETPLett2014}
{\sc \au{Dyachenko, S.A.}, \au{Lushnikov, P.M.} \& \au{Korotkevich, A.O.}}
  \yr{2014}  \at{{The complex singularity of a Stokes wave}}.  \jt{{JETP
  Letters}}  \bvol{98}~(11),  \pg{675--679}.

\bibitem[Dyachenko {\em et~al.\/}(2016)Dyachenko, Lushnikov \&
  Korotkevich]{dyachenko2016branch}
{\sc \au{Dyachenko, S.A.}, \au{Lushnikov, P.M.} \& \au{Korotkevich, A.O.}}
  \yr{2016}  \at{{Branch cuts of Stokes wave on deep water. Part I: numerical
  solution and Pad{\'e} approximation}}.  \jt{{Studies in Applied Mathematics}}
   \bvol{137}~(4),  \pg{419--472}.

\bibitem[Dyachenko \& Newell(2016)]{dyachenko2016whitecapping}
{\sc \au{Dyachenko, S.} \& \au{Newell, A.C.}} \yr{2016}  \at{Whitecapping}.
  \jt{{Studies in Applied Mathematics}}  \bvol{137}~(2),  \pg{199--213}.

\bibitem[Dyachenko \& Semenova(2023{\natexlab{{\em
  a\/}}})]{dyachenko2023canonical}
{\sc \au{Dyachenko, S.A.} \& \au{Semenova, A.}} \yr{2023{\natexlab{{\em a\/}}}}
   \at{{Canonical conformal variables based method for stability of Stokes
  waves}}.  \jt{{Studies in Applied Mathematics}}  \bvol{150}~(3),
  \pg{705--715}.

\bibitem[Dyachenko \& Semenova(2023{\natexlab{{\em
  b\/}}})]{dyachenko2023quasiperiodic}
{\sc \au{Dyachenko, S.A.} \& \au{Semenova, A.}} \yr{2023{\natexlab{{\em b\/}}}}
   \at{{Quasiperiodic perturbations of Stokes waves: Secondary bifurcations and
  stability}}.  \jt{Journal of Computational Physics}  \pg{p. 112411}.

\bibitem[Frigo \& Johnson(2005)]{frigo2005design}
{\sc \au{Frigo, M.} \& \au{Johnson, S.G.}} \yr{2005}  \at{{The design and
  implementation of FFTW3}}.  \jt{{Proceedings of the IEEE}}  \bvol{93}~(2),
  \pg{216--231}.

\bibitem[Grant(1973)]{grant1973singularity}
{\sc \au{Grant, M.A.}} \yr{1973}  \at{{The singularity at the crest of a finite
  amplitude progressive Stokes wave}}.  \jt{{Journal of Fluid Mechanics}}
  \bvol{59}~(part 2),  \pg{257--262}.

\bibitem[Haragus \& Kapitula(2008)]{haragucs2008spectra}
{\sc \au{Haragus, M.} \& \au{Kapitula, T.}} \yr{2008}  \at{{On the spectra of
  periodic waves for infinite-dimensional Hamiltonian systems}}.  \jt{{Physica
  D: Nonlinear Phenomena}}  \bvol{237}~(20),  \pg{2649--2671}.

\bibitem[Haziot {\em et~al.\/}(2022)Haziot, Hur, Strauss, Toland, Wahl{\'e}n,
  Walsh \& Wheeler]{haziotetcreview}
{\sc \au{Haziot, S.}, \au{Hur, V.}, \au{Strauss, W.}, \au{Toland, J.},
  \au{Wahl{\'e}n, E.}, \au{Walsh, S.} \& \au{Wheeler, M.}} \yr{2022}
  \at{Traveling water waves—the ebb and flow of two centuries}.
  \jt{Quarterly of applied mathematics}  \bvol{80},  \pg{317--401}.

\bibitem[Hur \& Yang(2023)]{hur2023unstable}
{\sc \au{Hur, V.M.} \& \au{Yang, Z.}} \yr{2023}  \at{{Unstable Stokes Waves}}.
  \jt{{Archive for Rational Mechanics and Analysis}}  \bvol{247}~(4),  \pg{62}.

\bibitem[Hur(2006)]{hur2006global}
{\sc \au{Hur, V.~M.}} \yr{2006}  \at{{Global Bifurcation Theory of Deep-Water
  Waves with Vorticity}}.  \jt{{SIAM Journal on Mathematical Analysis}}
  \bvol{37}~(5),  \pg{1482--1521}.

\bibitem[Kapitula \& Promislow(2013)]{kapitulapromislow}
{\sc \au{Kapitula, T.} \& \au{Promislow, K.}} \yr{2013} {\em {Spectral and
  dynamical stability of nonlinear waves}\/},  \st{{Applied Mathematical
  Sciences}},  \vol{vol. 185}.  \publ{Springer, New York}.

\bibitem[Korotkevich {\em et~al.\/}(2023)Korotkevich, Lushnikov, Semenova \&
  Dyachenko]{korotkevich2022superharmonic}
{\sc \au{Korotkevich, A.O.}, \au{Lushnikov, P.M.}, \au{Semenova, A.} \&
  \au{Dyachenko, S.A.}} \yr{2023}  \at{{Superharmonic instability of Stokes
  waves}}.  \jt{{Studies in Applied Mathematics}}  \bvol{150}~(1),
  \pg{119--134}.

\bibitem[Levi-Civita(1925)]{levi1925determination}
{\sc \au{Levi-Civita, T.}} \yr{1925}  \at{{D{\'e}termination rigoureuse des
  ondes permanentes d'ampleur finie}}.  \jt{{Mathematische Annalen}}
  \bvol{93}~(1),  \pg{264--314}.

\bibitem[Lighthill(1965)]{lighthill1965contributions}
{\sc \au{Lighthill, M.J.}} \yr{1965}  \at{{Contributions to the theory of waves
  in non-linear dispersive systems}}.  \jt{{IMA Journal of Applied
  Mathematics}}  \bvol{1}~(3),  \pg{269--306}.

\bibitem[Longuet-Higgins(1985)]{longuet1985bifurcation}
{\sc \au{Longuet-Higgins, MS}} \yr{1985}  \at{Bifurcation in gravity waves}.
  \jt{Journal of Fluid Mechanics}  \bvol{151},  \pg{457--475}.

\bibitem[Longuet-Higgins(2008)]{longuet2008approximation}
{\sc \au{Longuet-Higgins, M.S.}} \yr{2008}  \at{{On an approximation to the
  limiting Stokes wave in deep water}}.  \jt{{Wave Motion}}  \bvol{45}~(6),
  \pg{770--775}.

\bibitem[Longuet-Higgins \& Fox(1977)]{longuet1977theory}
{\sc \au{Longuet-Higgins, M.S.} \& \au{Fox, M.J.H.}} \yr{1977}  \at{{Theory of
  the almost-highest wave: the inner solution}}.  \jt{{Journal of Fluid
  Mechanics}}  \bvol{80}~(4),  \pg{721--741}.

\bibitem[Longuet-Higgins \& Fox(1978)]{longuet1978theory}
{\sc \au{Longuet-Higgins, M.S.} \& \au{Fox, M.J.H.}} \yr{1978}  \at{{Theory of
  the almost--highest wave. Part $2$. Matching and analytic extension}}.
  \jt{{Journal of Fluid Mechanics}}  \bvol{85},  \pg{769--786}.

\bibitem[Longuet-Higgins \& Tanaka(1997)]{longuet1997crest}
{\sc \au{Longuet-Higgins, M.S.} \& \au{Tanaka, M.}} \yr{1997}  \at{{On the
  crest instabilities of steep surface waves}}.  \jt{Journal of Fluid
  Mechanics}  \bvol{336},  \pg{51--68}.

\bibitem[Longuet-Higgins(1975)]{longuet1975integral}
{\sc \au{Longuet-Higgins, M.~S.}} \yr{1975}  \at{{Integral properties of
  periodic gravity waves of finite amplitude}}.  \jt{{Proceedings of the Royal
  Society of London. A. Mathematical and Physical Sciences}}
  \bvol{342}~(1629),  \pg{157--174}.

\bibitem[Lushnikov {\em et~al.\/}(2017)Lushnikov, Dyachenko \&
  Silantyev]{lushnikov2017new}
{\sc \au{Lushnikov, P.M.}, \au{Dyachenko, S.A.} \& \au{Silantyev, D.A.}}
  \yr{2017}  \at{{New conformal mapping for adaptive resolving of the complex
  singularities of Stokes wave}}.  \jt{{Proceedings of the Royal Society A:
  Mathematical, Physical and Engineering Sciences}}  \bvol{473}~(2202),
  \pg{20170198}.

\bibitem[MacKay \& Saffman(1986)]{mackay1986stability}
{\sc \au{MacKay, R.~S.} \& \au{Saffman, P.~G.}} \yr{1986}  \at{{Stability of
  water waves}}.  \jt{{Proceedings of the Royal Society of London. A.
  Mathematical and Physical Sciences}}  \bvol{406}~(1830),  \pg{115--125}.

\bibitem[Maklakov(2002)]{maklakov2002almost}
{\sc \au{Maklakov, D.V.}} \yr{2002}  \at{{Almost-highest gravity waves on water
  of finite depth}}.  \jt{{European Journal of Applied Mathematics}}
  \bvol{13}~(1),  \pg{67}.

\bibitem[Michell(1893)]{michell1893}
{\sc \au{Michell, J.H.}} \yr{1893}  \at{{XLIV. The highest waves in water}}.
  \jt{{The London, Edinburgh, and Dublin Philosophical Magazine and Journal of
  Science}}  \bvol{36}~(222),  \pg{430--437}.

\bibitem[Nekrasov(1921)]{nekrasov1921waves}
{\sc \au{Nekrasov, A.I.}} \yr{1921}  \at{{On waves of permanent type I}}.
  \jt{Izv. Ivanovo-Voznesensk. Polite. Inst.}  \bvol{3},  \pg{52--65}.

\bibitem[Nguyen \& Strauss(2023)]{nguyenstrauss}
{\sc \au{Nguyen, H.Q.} \& \au{Strauss, W.A.}} \yr{2023}  \at{{Proof of
  Modulational Instability of Stokes Waves in Deep Water}}.
  \jt{{Communications on Pure and Applied Mathematics}}  \bvol{76}~(5),
  \pg{1035--1084}.

\bibitem[Onorato {\em et~al.\/}(2013)Onorato, Residori, Bortolozzo, Montina \&
  Arecchi]{onorato2013rogue}
{\sc \au{Onorato, M.}, \au{Residori, S.}, \au{Bortolozzo, U.}, \au{Montina, A.}
  \& \au{Arecchi, F.~T.}} \yr{2013}  \at{Rogue waves and their generating
  mechanisms in different physical contexts}.  \jt{Physics Reports}
  \bvol{528}~(2),  \pg{47--89}.

\bibitem[Ovsyannikov(1973)]{ovsyannikov1973dynamika}
{\sc \au{Ovsyannikov, L.V.}} \yr{1973}  \at{{Dynamika sploshnoi sredy,
  Lavrentiev Institute of Hydrodynamics}}.  \jt{Sib. Branch Acad. Sci. USSR}
  \bvol{15},  \pg{104}.

\bibitem[Plotnikov(1982)]{plotnikov1982}
{\sc \au{Plotnikov, P.I.}} \yr{1982}  \at{{Justification of the Stokes
  conjecture in the theory of surface waves (in Russian)}}.  \jt{{Dinamika
  Sploshnoi Sredy}}  \bvol{57},  \pg{4176}.

\bibitem[Plotnikov(2002)]{plotnikov2002proof}
{\sc \au{Plotnikov, P.I.}} \yr{2002}  \at{{A proof of the Stokes conjecture in
  the theory of surface waves}}.  \jt{Studies in Applied Mathematics}
  \bvol{108}~(2),  \pg{217--244}.

\bibitem[Saffman(1985)]{saffman1985superharmonic}
{\sc \au{Saffman, P.~G.}} \yr{1985}  \at{{The superharmonic instability of
  finite-amplitude water waves}}.  \jt{{Journal of Fluid Mechanics}}
  \bvol{159},  \pg{169--174}.

\bibitem[Schwartz(1974)]{schwartz1974computer}
{\sc \au{Schwartz, L.W.}} \yr{1974}  \at{{Computer extension and analytic
  continuation of Stokes’ expansion for gravity waves}}.  \jt{{Journal of
  Fluid Mechanics}}  \bvol{62}~(3),  \pg{553--578}.

\bibitem[Silantyev(2019)]{dennisnew}
{\sc \au{Silantyev, D.~A.}} \yr{2019}  \at{{A new conformal map for computing
  Stokes wave}}.  \jt{private communications} .

\bibitem[Stewart(2002)]{stewart2002krylov}
{\sc \au{Stewart, G.W.}} \yr{2002}  \at{{A Krylov-Schur algorithm for large
  eigenproblems}}.  \jt{{SIAM Journal on Matrix Analysis and Applications}}
  \bvol{23}~(3),  \pg{601--614}.

\bibitem[Stokes(1847)]{stokes1847theory}
{\sc \au{Stokes, G.G.}} \yr{1847}  \at{{On the theory of oscillatory waves}}.
  \jt{{Transactions of the Cambridge Philosophical Society}}  \bvol{8},
  \pg{441}.

\bibitem[Stokes(1880{\natexlab{{\em a\/}}})]{stokes1880theory}
{\sc \au{Stokes, G.G.}} \yr{1880{\natexlab{{\em a\/}}}}  \at{{On the theory of
  oscillatory waves}}.  \jt{Mathematical and Physical Papers}  \bvol{1},
  \pg{197}.

\bibitem[Stokes(1880{\natexlab{{\em b\/}}})]{stokes1880}
{\sc \au{Stokes, G.G.}} \yr{1880{\natexlab{{\em b\/}}}}  \at{{Supplement to a
  paper on the Theory of Oscillatory Waves}}.  \jt{{Mathematical and Physical
  Papers}}  \bvol{1},  \pg{314}.

\bibitem[Tanaka(1983)]{tanaka1983stability}
{\sc \au{Tanaka, M.}} \yr{1983}  \at{{The stability of steep gravity waves}}.
  \jt{{Journal of the Physical Society of Japan}}  \bvol{52}~(9),
  \pg{3047--3055}.

\bibitem[Tanaka(1985)]{tanaka1985stability}
{\sc \au{Tanaka, M.}} \yr{1985}  \at{{The stability of steep gravity waves.
  Part 2}}.  \jt{{Journal of Fluid Mechanics}}  \bvol{156},  \pg{281--289}.

\bibitem[Tanveer(1991)]{tanveer1991singularities}
{\sc \au{Tanveer, S.}} \yr{1991}  \at{{Singularities in water waves and
  Rayleigh--Taylor instability}}.  \jt{{Proceedings of the Royal Society of
  London. Series A: Mathematical and Physical Sciences}}  \bvol{435}~(1893),
  \pg{137--158}.

\bibitem[Tanveer(1993)]{tanveer1993singularities}
{\sc \au{Tanveer, S.}} \yr{1993}  \at{Singularities in the classical
  {Rayleigh-Taylor} flow: formation and subsequent motion}.  \jt{{Proceedings
  of the Royal Society of London. Series A: Mathematical and Physical
  Sciences}}  \bvol{441}~(1913),  \pg{501--525}.

\bibitem[Toland(1978)]{toland1978existence}
{\sc \au{Toland, J.F.}} \yr{1978}  \at{{On the existence of a wave of greatest
  height and Stokes’s conjecture}}.  \jt{{Proceedings of the Royal Society of
  London. A. Mathematical and Physical Sciences}}  \bvol{363}~(1715),
  \pg{469--485}.

\bibitem[Toland(1996)]{toland1996stokes}
{\sc \au{Toland, J.~F.}} \yr{1996}  \at{{Stokes waves}}.  \jt{{Topological
  Methods in Nonlinear Aanalysis}}  \bvol{7}~(1),  \pg{1--48}.

\bibitem[Whitham(1967)]{whitham1967non}
{\sc \au{Whitham, G.B.}} \yr{1967}  \at{{Non-linear dispersion of water
  waves}}.  \jt{{Journal of Fluid Mechanics}}  \bvol{27}~(2),  \pg{399--412}.

\bibitem[Williams(1981)]{williams1981limiting}
{\sc \au{Williams, J.M.}} \yr{1981}  \at{{Limiting gravity waves in water of
  finite depth}}.  \jt{{Philosophical Transactions of the Royal Society of
  London. Series A, Mathematical and Physical Sciences}}  \bvol{302}~(1466),
  \pg{139--188}.

\bibitem[Williams(1985)]{williams1985tables}
{\sc \au{Williams, J.M.}} \yr{1985} {\em {Tables of progressive gravity
  waves}\/}.  \publ{{Boston : Pitman Advanced Pub. Program}}.

\bibitem[Zakharov {\em et~al.\/}(1996)Zakharov, Kuznetsov \&
  Dyachenko]{dyachenko1996dynamics}
{\sc \au{Zakharov, V.E.}, \au{Kuznetsov, E.A.} \& \au{Dyachenko, A.I.}}
  \yr{1996}  \at{{Dynamics of free surface of an ideal fluid without gravity
  and surface tension}}.  \jt{{Fizika Plasmy}}  \bvol{22},  \pg{916--928}.

\bibitem[Zakharov(1968)]{zakharov1968}
{\sc \au{Zakharov, V.~E.}} \yr{1968}  \at{{Stability of periodic waves of
  finite amplitude on a surface}}.  \jt{{Journal of Applied Mechanics and
  Technical Physics}}  \bvol{9}~(2),  \pg{190--194}.

\bibitem[Zufiria(1987)]{zufiria1987non}
{\sc \au{Zufiria, J.~A.}} \yr{1987}  \at{Non-symmetric gravity waves on water
  of infinite depth}.  \jt{Journal of Fluid Mechanics}  \bvol{181},
  \pg{17--39}.

\end{thebibliography}

\end{document}